\documentclass[twocolumn,prl,showpacs]{revtex4}
\usepackage{multirow}
\usepackage{graphicx}
\usepackage{amsmath}
\usepackage{enumerate}
\usepackage{epstopdf}
\usepackage{times}
\usepackage{color}
\usepackage{float}

\usepackage[colorlinks,bookmarks=false,citecolor=blue,linkcolor=red,urlcolor=blue]{hyperref}

\begin{document}
\title{
Half-metallic magnetization plateaux}
\author{Zhihao Hao}
\affiliation{Department of Physics and Astronomy, University of Waterloo, Waterloo, ON N2L 3G1, Canada}
\author{Oleg A. Starykh}
\affiliation{Department of Physics and Astronomy, University of Utah, Salt Lake City, UT 84112, USA}
\begin{abstract}
We propose a novel interaction-based route to half-metal state for interacting electrons on two-dimensional lattices.
Magnetic field applied parallel to the lattice is used to tune one of the spin densities to a particular commensurate with the lattice
value in which the system spontaneously `locks in' via van Hove enhanced density wave state. Electrons of opposite
spin polarization retain their metallic character and provide for the half-metal state which, in addition, supports magnetization
plateau in a finite interval of external magnetic field. Similar half-metal state is realized in the finite-U Hubbard model on a triangular 
lattice at $1/3$ of the maximum magnetization.
\end{abstract}
\date{July 30, 2012}
\maketitle


Spin systems supporting robust magnetization plateaux whereby macroscopic magnetization $M$ is fixed
at a rational fraction of the full (saturated) magnetization value $M_{\rm sat}$ in a finite interval of external
magnetic field $h_1 < h < h_2$ are subject of intense experimental studies \cite{kodama2002,azurite2005,ueda2005,fortune2009,nema2009}.
Typically these materials are magnetic insulators which are well described by the Heisenberg-type models
with short-range exchange interactions between localized spins.

One of the best understood and studied plateau states is represented by the up-up-down (UUD) magnetization plateau
at $M = \frac{1}{3} M_{\rm sat}$
in the triangular lattice antiferromagnet \cite{kawamura1985,chubukov1991}. This 1/3 magnetization plateau
is a remarkably stable state -- it is known to survive significant spatial deformation of exchange integrals
in both quantum (spin 1/2) and classical versions of the model \cite{alicea2009,griset2011}, well beyond the point where adjacent to it
co-planar spin states ceases to exist altogether.
The basic reason for this stability lies in the {\em collinear} structure of the UUD configuration. Collinearity preserves
U$(1)$ symmetry of the Hamiltonian with respect to the magnetic field axis. The only symmetry that the UUD state
breaks is the discrete translational symmetry -- its unit cell consists of two up and one down spin.
This ensures the absence of the gapless (Goldstone) modes in the spectrum and
implies enhanced stability of the collinear spin arrangement.

Since the Heisenberg model is just a low-energy approximation to the large-$U/t$ limit (here $t$ is the
hopping integral and $U$ is on-site interaction energy, see below) of the half-filled Hubbard model,
the \textit{insulating} magnetization plateau state is favored by strong electron-electron interactions.
What happens to the $1/3$ magnetization plateau state as the ratio $U/t$ is reduced
and electrons delocalize is one of the key questions motivating our study.

A different class of magnetization plateau materials is provided by half-metallic ferromagnets in
which by virtue of peculiar electronic structure {\em all} conduction electrons have the same spin orientation (say, up, for definiteness).
In their simplest version half-metallic materials are then fully saturated, $M = M_{\rm sat}$.
As the name suggests, these materials are conductors and are well understood in terms of
non-interacting electron picture \cite{degroot1983,katsnelson2008}.

The aim of our work is to unite these phenomena by
proposing two new {\em interacting} routes to the {\em half-metallic} magnetization plateau states.
Both routes require finite external (Zeeman) magnetic field, applied parallel to the two-dimensional triangular lattice.

\begin{figure}
\centering
\includegraphics[width=0.8\columnwidth]{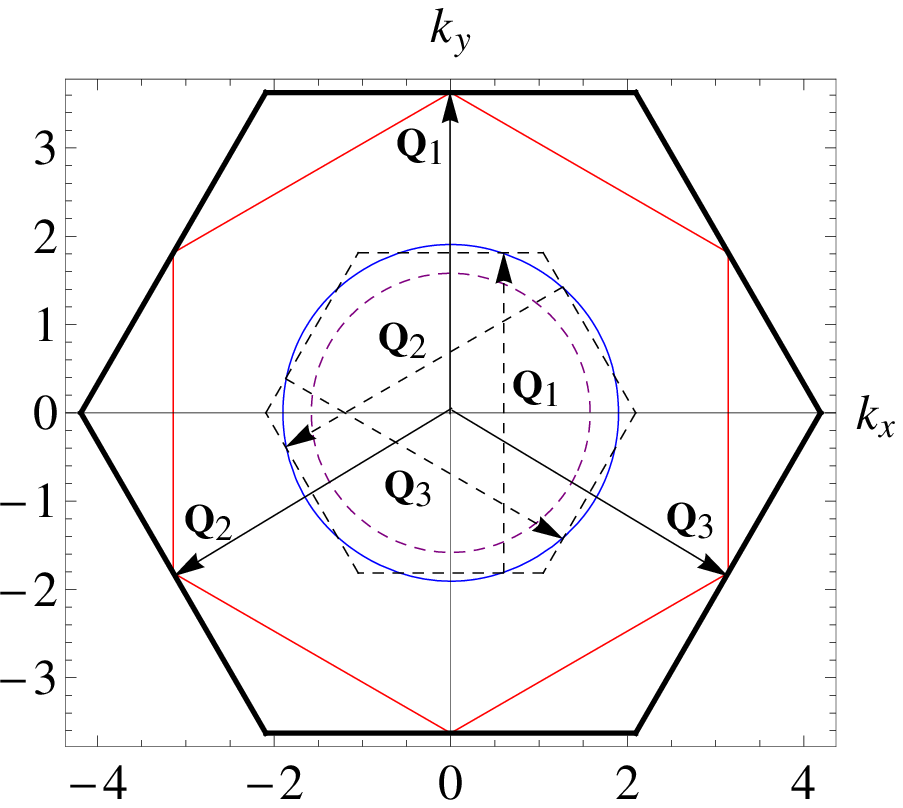}\\
\includegraphics[width=0.8\columnwidth]{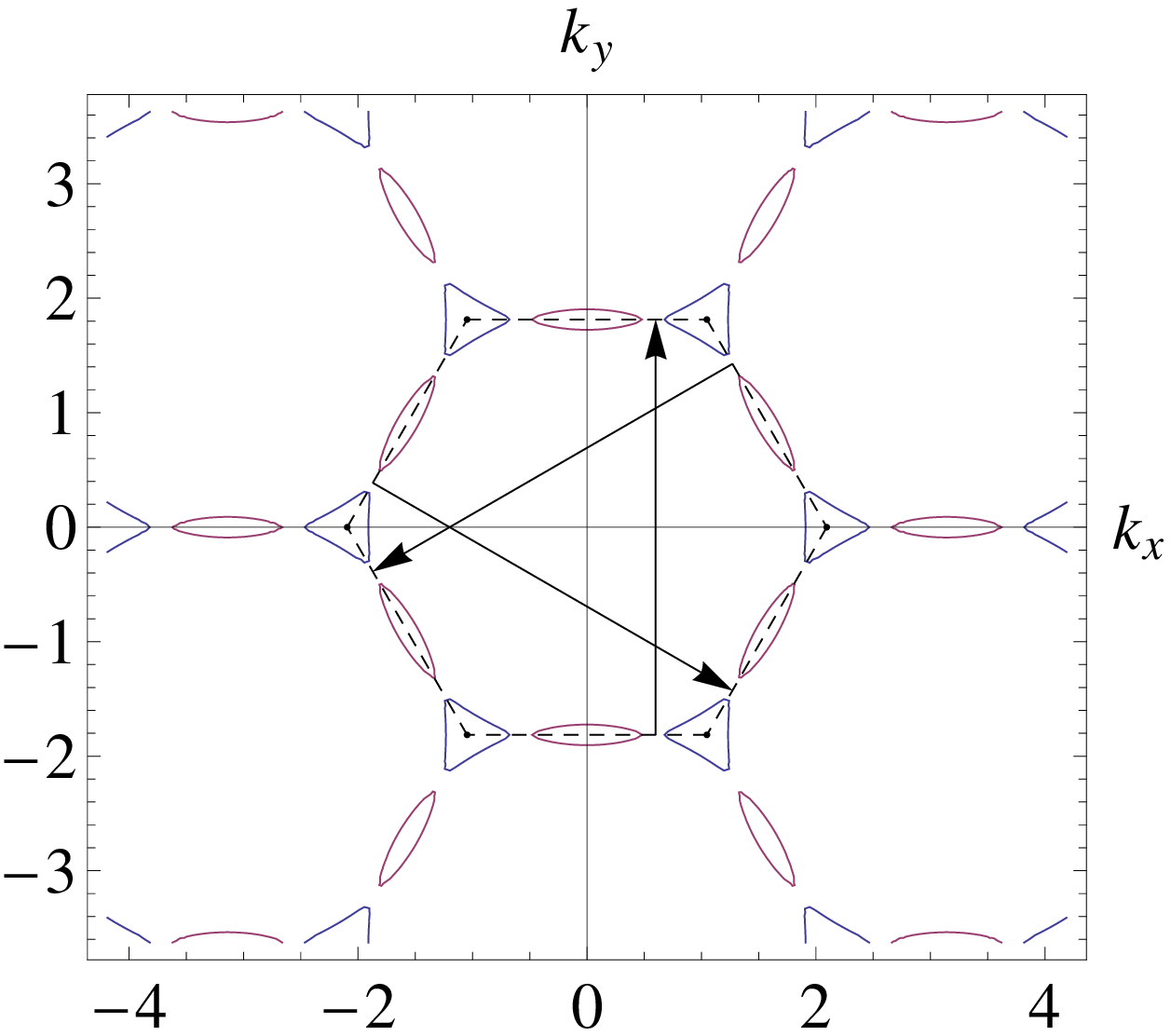}\\
\caption{(Color online) The non-interacting Fermi surfaces of spin-up and spin-down electrons (top) at $1/2$ magnetization and
the reconstructed fermi surface of spin-down electrons at $y_\downarrow/t = 0.1$ (bottom).
The thick black hexagon is the first Brillouin zone of triangular lattice. The red hexagon is the Fermi surface of the spin-up electron.
It is perfectly nested by linear combinations of three wave vectors (the arrows). The dashed hexagon is the first Brillouin zone under
the folded scheme. The dashed purple circle is an example of the hot-spot-free Fermi surface of spin-down electrons in the hole doped system.}
\label{fig:rec}
\end{figure}

The {\em weak-coupling} route, described first below, relies on tuning density of majority (say, spin-up) electrons $n_\uparrow$ to a specific value ($3/4$),
commensurate with the triangular lattice, at which the Fermi surface (FS) passes via a set of van Hove
points with logarithmically divergent density of states, see Figure~\ref{fig:rec}. Depending on the total electron density $n = n_\uparrow + n_\downarrow$,
the FS of minority (spin-down) electrons may or may not be affected by the interactions, but in any case retains its metallic character.
The resulting ground state is half-metal which supports $M=(\frac{3}{4} - n_\downarrow) M_{\rm sat}$ magnetization
plateau with ferrimagnetic (up-down-down-down) collinear spin structure.
We emphasize that $M/M_{\rm stat}$ is generally {\em irrational}. This novel state has no analogs in the
in the large-$U$ limit of the Hubbard model and can be pictured as a phase with coexisting {\em spin-} and {\em charge-}density wave orders.
Theoretical analysis of this limit bears strong similarities with recent proposals \cite{martin2008,nandkishore2012,chern2012,batista2012}
of collinear and chiral spin-density wave (SDW) states of itinerant electrons on honeycomb lattice in vicinity of
electron filling factors $n=3/8$ and $5/8$ and at zero magnetization. In contrast to our problem, however, SDW order there spontaneously breaks
spin-rotational symmetry of the Hamiltonian and, as a result, is accompanied by gapless collective excitations \cite{metlitski2010}
which drive the competition between the collinear and chiral orders at finite temperature \cite{nandkishore2012}.
This complication is absent in our problem where external magnetic field sets direction of the collinear SDW.
The resulting half-metallic state breaks only discrete translational symmetry of the lattice and is stable
with respect to fluctuations of the order parameter about its mean-field value.

Next we describe the {\em strong coupling} (large-$U$) route and show that $M=\frac{1}{3} M_{\rm sat}$ magnetization plateau
($n_\uparrow = 2/3, n_\downarrow = 1/3$), present in the $U/t \to \infty$ limit,
survives down to $U_{c1}/t \approx 4.3$, which is significantly lower than the zero-magnetization critical value
$U_{120^\circ}/t \approx 10$ below which the three-sublattice $120^\circ$ magnetic order melts as the system
transitions to a quantum spin-liquid state \cite{morita2002,sahebsara2008,lauchli2010}.
Quite interestingly, we find that in the interval $U_{c1} \leq U \leq U_{c2} \approx 4.8 t$
the UUD state is a {\em half-metal} with mobile majority (spin-up) electrons. A
closely related question of the evolution of the ground state of the Hubbard model at {\em zero} magnetization, as a function of $U/t$,
has been actively investigated \cite{morita2002,singh2005,sahebsara2008,davoudi2008,lauchli2010}, partly because of the potential relevance
to intriguing `spin-liquid' organic insulators \cite{motrunich2005,lee2005,kanoda2011}.

{\em Weak-coupling analysis} is based on the extended Hubbard model description of electrons on  triangular lattice:
\begin{eqnarray}
\label{eqn:hami}
H&=&-t\sum_{\langle r r'\rangle}(c_{r\sigma}^\dagger c_{r'\sigma}+{\rm h.c})+U\sum_{r} n_{r\uparrow}n_{r\downarrow}+V\sum_{\langle r r'\rangle} n_{r}n_{r'}\nonumber\\
&&-\sum_{r}\left(\mu+\frac{h\sigma}{2}\right)c_{r\sigma}^\dagger c_{r\sigma}.
\end{eqnarray}
Here $\langle r r'\rangle$ labels nearest neighbour bonds, $0 < U/t,V/t \ll 1$ are onsite and nearest neighbour repulsive interactions, respectively.
For now we set the filling factor at $n=\langle n_i \rangle = \langle n_{i\uparrow} + n_{i\downarrow}\rangle = 1$.

The Zeeman field $h$, normalized to include the usual $g \mu_B$ factor, is tuned to produce $M = \frac{1}{2} M_{\rm sat}$ magnetization so that on
average $n_\uparrow = 3/4$ and $n_\downarrow = 1/4$ per site. Under this condition the
Fermi surface of spin-up electrons, shown in Figure~\ref{fig:rec}, is given by a perfect hexagon whose vertices, located at the M points of the first Brillouin zone (BZ),
are van Hove (saddle) points of the dispersion with vanishing Fermi velocity \cite{nandkishore2012}.
These saddle points are the reason for the (logarithmically) diverging density of states as well as singular susceptibility (see below).
They are connected by the wave vectors
$\mathbf{Q}_1=\frac{2\pi}{\sqrt{3}}\hat{y}$ and $\mathbf{Q}_{2,3}=\mp \pi\hat{x}-\frac{\pi}{\sqrt{3}}\hat{y}$ (Fig. \ref{fig:rec}),
which are just halfs of the corresponding reciprocal lattice vectors $\mathbf{G}_{1,2,3}$ \cite{martin2008}.
In addition, parallel faces of the FS are perfectly nested by linear combinations of $\mathbf{Q}$'s.
Spin-down FS is nearly circular, as Fig. \ref{fig:rec} shows, and does not possess any special features.

The very special role of the wave vectors $\mathbf{Q}_{1,2,3}$ is conveniently quantified
by charge susceptibility $\chi_{\sigma}(\mathbf{q})$ of spin-$\sigma$ electrons defined in the standard way as
\begin{equation}
\label{eqn:sus}
    \chi_{\sigma}(\mathbf{q})=\frac{1}{N}
    \sum_{\mathbf{k}}\frac{n_{\mathbf{k},\sigma}-n_{\mathbf{k}+\mathbf{q},\sigma}}
    {\epsilon_{\mathbf{k}}-\epsilon_{\mathbf{k}+\mathbf{q}}}.
\end{equation}
Here $N$ is the number of sites, $n_{\mathbf{k},\sigma}$ is the occupation number of fermions with spin $\sigma$ and momentum $\mathbf{k}$ and
\begin{equation}\label{eqn:ek}
\epsilon_{\mathbf{k}}=-2t\Big(\cos k_x+2\cos\frac{k_x}{2}\cos\frac{\sqrt{3}k_y}{2}\Big)
\end{equation}
is the free-particle dispersion. Straightforward calculation, outlined in \cite{suppl},
shows that susceptibility of spin-up electrons
is strongly divergent at $\mathbf{q} = \pm \mathbf{Q}_a$, while that of spin-down electrons is finite,
\begin{eqnarray}
\label{eq:sus-up}
\chi_{\uparrow}(\mathbf{Q}_a) &=& -\frac{C_\uparrow}{t} \ln^2\Big(\frac{\Lambda}{q_0}\Big), ~C_\uparrow = \frac{1}{2\pi^2}, \\
\chi_{\downarrow}(\mathbf{Q}_a) &=& -\frac{C_\downarrow}{t}, ~C_\downarrow \approx0.136 .
\label{eq:sus-down}
\end{eqnarray}
Here $\Lambda\sim\pi/a$ is the `size' of the BZ ($a$ is the lattice spacing)
while $q_0 \sim 1/L$ is the microscopic cut-off which scales as the inverse linear size of the lattice $L \sim \sqrt{N} a$.

Eq.~\eqref{eq:sus-up} suggests strong modulation of density at $\mathbf{q} = \pm \mathbf{Q}_a$ which motivates
the following mean-field ansatz
\begin{equation}
\label{eq:ansatz}
\langle n_{r, \sigma}\rangle = \frac{2+\sigma}{4} + m_\sigma\sum_{a=1}^{3}\cos(\mathbf{Q}_a \cdot \mathbf{r}) ,
\end{equation}
where index $\sigma$ describes two spin projections, $\sigma= \uparrow = +1$ and $\sigma= \downarrow = -1$, in the usual way.
Note that the filling factors $n_\sigma = \sum_{\mathbf{r}} \langle n_{r, \sigma}\rangle /N$ are
not affected by finite order parameters $m_\sigma$, and $n_\uparrow = 3/4$, while $n_\downarrow = 1/4$.
Ansatz \eqref{eq:ansatz} allows us to approximate \eqref{eqn:hami} as
\begin{eqnarray}
\label{eq:ham-mf}
H&=&\sum_{\mathbf{k}, \sigma} (\epsilon_{\mathbf{k}} - \mu_\sigma) c^+_{\mathbf{k} \sigma} c_{\mathbf{k} \sigma} +
\frac{1}{2} \sum_{\mathbf{k}, \sigma, a} y_\sigma (c^+_{\mathbf{k} \sigma} c_{\mathbf{k}+\mathbf{Q}_a \sigma}  + {\rm h.c.}) \nonumber\\
&& + 3 N V (m_\uparrow + m_\downarrow)^2 - 3 N U m_\uparrow m_\downarrow ,\\
&&y_\sigma = U m_{-\sigma} -  2V (m_\uparrow + m_\downarrow) ,
\label{eq:y}
\end{eqnarray}
The amplitudes of density modulation of spin-$\sigma$ electrons are determined self-consistently
\begin{eqnarray}
m_\sigma &=& \frac{1}{3 N} \sum_{\mathbf{r}, a} \cos(\mathbf{Q}_a \cdot \mathbf{r}) \langle n_{\mathbf{r}, \sigma} \rangle \nonumber\\
&&=\frac{y_\sigma}{3 N} \sum_{\mathbf{k}, a} \frac{n_{\mathbf{k},\sigma}-n_{\mathbf{k}+\mathbf{Q}_a,\sigma}}
    {\tilde\epsilon_{\mathbf{k}}-\tilde\epsilon_{\mathbf{k}+\mathbf{Q}_a}} = y_\sigma \chi_\sigma(\mathbf{Q}_a) ,
\label{eq:m-sigma}
\end{eqnarray}
where dispersion $\tilde\epsilon_{\mathbf{k}}$ is determined by the mean-field Hamiltonian \eqref{eq:ham-mf}.
We find $m_\downarrow = C_\downarrow(V-U/2) m_\uparrow/t$, which means $|m_\downarrow| \ll |m_\uparrow|$, and
\begin{equation}
\frac{C_\uparrow}{t} \ln^2\Big(\frac{\Lambda}{q_\uparrow}\Big) = \Big(V + \frac{C_\downarrow}{t}(V - \frac{U}{2})^2\Big)^{-1} .
\end{equation}
The main effect of interaction $y_\uparrow$ in \eqref{eq:ham-mf} is to provide an infra-red cut-off $q_\uparrow \sim |y_\uparrow/t|^{1/2}$
in the susceptibility \eqref{eq:sus-up} of spin-up electrons at the van Hove points.
This leads to the final result
\begin{eqnarray}
\label{eq:m}
m_\uparrow &=& -\frac{\Lambda^2 t}{V+\frac{C_\downarrow}{t}(V - U/2)^2} \nonumber\\
&&\times \exp\left(-\sqrt{\frac{4 t}{C_\uparrow [V+\frac{C_\downarrow}{t}(V - U/2)^2]}}\right).
\end{eqnarray}
Non-analytic dependence of $m$ on interaction amplitudes $U, V$ is determined by van Hove points.
The sign of $m_\uparrow$ is chosen such that the FS of spin-up electrons is gapped for {\em all} momenta.
We checked that the opposite sign leads to a quadratic touching of the top two bands at $\Gamma$ point and results in a state of higher energy \cite{suppl}.

Equations \eqref{eq:ansatz} and \eqref{eq:m} show that the ground state is a superposition of commensurate
charge-density and collinear spin-density waves. This is a four-sublattice state: observe that
$\sum_{a=1}^{3}\cos(\mathbf{Q}_a \cdot \mathbf{r})$ takes values $3, -1, -1, -1$ on the sites of triangular lattice
$\mathbf{r} = (x, y) = d_1 \mathbf{a}_1 + d_2 \mathbf{a}_2$, where $\mathbf{a}_1 = (1,0)$ and $\mathbf{a}_2 = (1/2, \sqrt{3}/2)$
are elementary lattice vectors and $d_{1,2}$ are integers.
We stress that the perfectly nested FS of spin-up electrons is crucial for the spin-up electrons to be gapped at arbitrary weak interaction.

Note that in the absence of direct density-density interaction, when $V=0$, the order parameter
scales as $m\sim \exp\left(-{\rm const}/U\right)$. The density wave of spin-up electrons in this case is driven by the
effective interaction $\propto \chi_\downarrow U^2$, which is mediated by spin-down electrons,
since the onsite repulsion $U$ only couples electrons of opposite spins.
A small finite $V \geq U^2/t$, see  \eqref{eq:m}, changes this scaling to a much stronger dependence $m\sim \exp\left(-{\rm const}/\sqrt{V}\right)$.

The band structure of spin-down electrons is also modified as \eqref{eq:ham-mf} shows.
For finite $y_\downarrow$, while the spin-down electrons remain gapless, the `hot spots' on the spin-down FS
(which are the points connected by $\mathbf{Q}_a$) are gapped and the FS is reconstructed as shown in Figure \ref{fig:rec}.
By reducing the density of spin-down electrons $n_\downarrow$ below $1/4$,
while maintaining that of spin-up ones at the perfect nesting condition $n_\uparrow = 3/4$, one
can reduce the spin-down FS below the critical volume to make it fit inside the reduced Brillouin zone
(dashed hexagon in Fig. \ref{fig:rec}), which is 4 times smaller than the original one,
and the hot spots disappear altogether.
We find that this happens for $n_{\rm cr} < 0.976$ (Figure \ref{fig:rec}), {\em i.e.} $n_{{\rm cr} \downarrow} < 0.226$.
Under this condition, and for weak interactions $U, V \ll t$, the FS of spin-down electrons is not affected
by the reconstruction of the FS of spin-up electrons at all. In either case, the result is a {\em half-metal}  where {\em all}
conducting electrons have spin opposite to the direction of the external field.

Similar half-metallic state can be realized on other lattices. For example, consider the same Hamiltonian \eqref{eqn:hami}
on the square lattice with filling factor $n\neq1$. A proper field $h$ can be applied such that the FS of the spin-up (if total density $n<1$) or
the spin-down (if $n>1$) electrons is perfectly nested. Clearly this FS is a square with vertices at $(\pm \pi, 0)$ and $(0, \pm \pi)$ points.
The system then develops density wave order at wave vector $\mathbf{Q}=(\pi,\pi)$.
The ordered state is generically half-metallic. Note also that in this case the FS of conduction electrons experiences no reconstruction
due to the absence of hot spots.

\begin{figure}
\centering
\includegraphics[width=0.9\columnwidth]{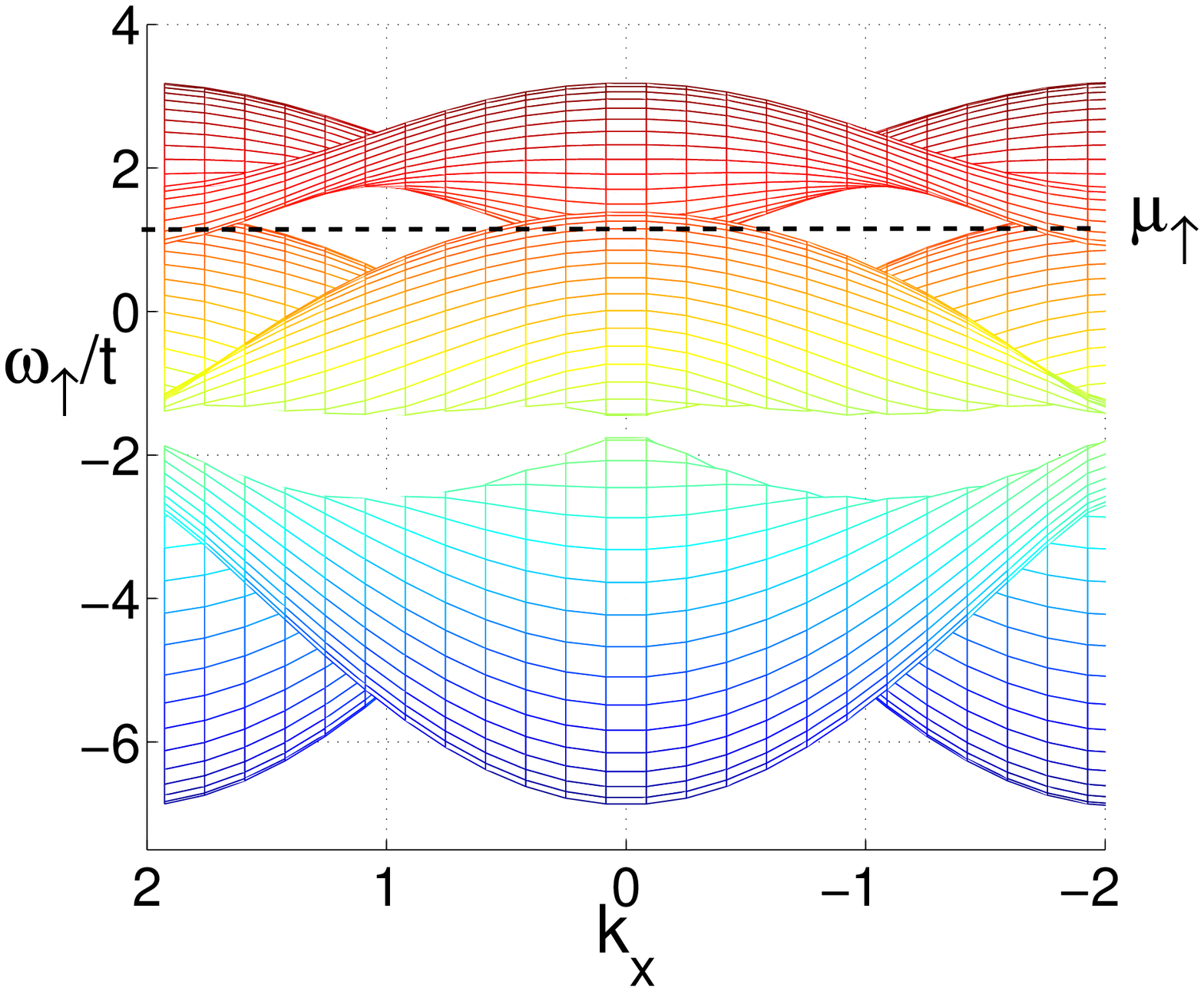}\\
\includegraphics[width=0.9\columnwidth]{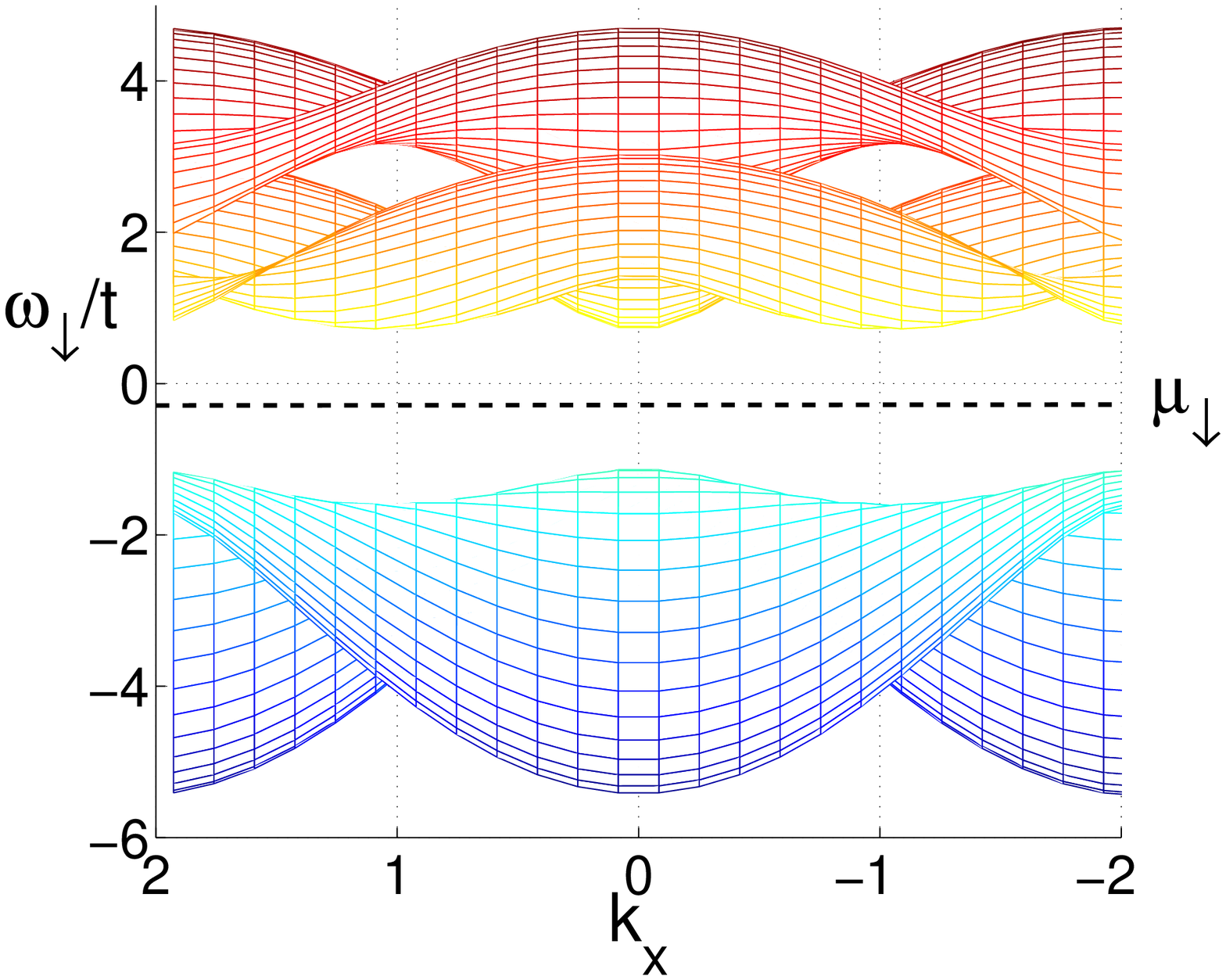}\\
\caption{(Color online) The spin-up (top) and spin-down (bottom) bands of the half-metal state at $1/3$ magnetization at $U=4.60t$.
In order to give an idea of dispersion along the $k_y$ axis as well, the graph shows
projections of two-dimensional bands onto $k_x$ axis for 30 different discrete $k_y$ values.
The Fermi energies $\mu_\sigma = \mu + \sigma (h/2 + U/6)$ are shown as black-dash lines. }
\label{fig:bands}
\end{figure}

{\em Strong coupling limit.} We now turn to the question of $M=\frac{1}{3} M_{\rm sat}$ plateau which exists in the opposite limit of
strong interactions, $U \gg t$.  As should be clear from the previous discussion, in the current situation filling factors of spin-up
(down) electrons $2/3 (1/3)$ do not correspond to perfectly nested FS.
In order to make connection with the insulating magnetization plateau phase of the Heisenberg spin model we set $V=0$,
keep total density at one electron per site $n = 1$ and introduce the following mean-field ansatz:
\begin{equation}
\label{eqn:mf13}
\langle n_{\mathbf{r}\sigma}\rangle=\frac{1}{2}+\frac{\sigma}{6}-2\eta_{\sigma}\cos(\mathbf{Q}\cdot\mathbf{r})
\end{equation}
where $\mathbf{Q}=\frac{4\pi}{3}\hat{x}$ describes the UUD pattern.
(Note that $\cos(\mathbf{Q}\cdot\mathbf{r})$ takes values $1, -1/2, -1/2$ on the triangular lattice.) Parameters $\eta_{\sigma}$ of the state are
determined self-consistently by the equations similar to \eqref{eq:m-sigma}. Solving them numerically we find discontinuous jump of
$\eta_\sigma$ from zero to finite values when $U\ge 4.30t$ for a range of $h$. We also find that $\eta_{\uparrow}$ and $\eta_{\downarrow}$ are
in general different so that the system displays a co-existence of the spin-density and charge-density wave orders \cite{suppl}.
Interestingly, it is now the spin-down electrons that are gapped while the spin-up electrons remain gapless around the
Fermi energy, see Figure \ref{fig:bands}. Spin-down electrons fill completely the lowest of the three bands
$\omega_\downarrow(\mathbf{k})$ in the folded Brillouin zone,
while the spin-up ones fill the two lowest bands $\omega_\uparrow(\mathbf{k})$.
For even stronger interaction $U> 4.80t$, the two upper spin-up bands also get separated
by a gap which turns the half-metal state into an insulator with collinear UUD pattern of local magnetization.

\begin{figure}
\centering
\includegraphics[width=0.9\columnwidth]{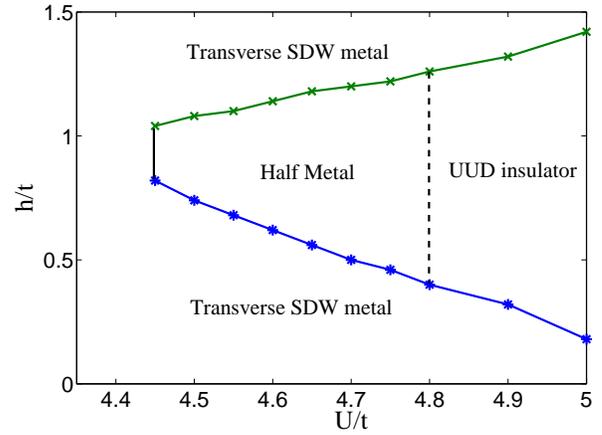}\\
\caption{(Color online) The mean-field phase boundaries of the half-metal and insulating states at $M = 1/3 ~M_{\rm sat}$ on triangular lattice.
Solid (dashed) lines denote first (second) order transitions. }
\label{fig:phase}
\end{figure}

To estimate how competitive the found half-metal state is we compared its energy with the uniformly magnetized transverse
spin-density wave state, {\em i.e.} ``cone'' state in magnetic language.
This state is characterized by the two order parameters, longitudinal $\langle S^z_\mathbf{r} \rangle = M$ and
transverse $\langle S^+_{\mathbf{q}}\rangle = m_0 e^{i \mathbf{q}\cdot \mathbf{r}}$ magnetizations \cite{krishnamurthy1990,jayaprakash1991},
where the ordering wave vector $\mathbf{q}$ is incommensurate in general.
Comparing their mean-field energies, we determined that the half-metal state has a lower energy for $4.45t\le U\le 4.80 t$ \cite{suppl}.
The resulting mean-field phase diagram is shown in Figure \ref{fig:phase}.
Note that both half-metal and insulator phases are magnetization plateau states with $M=\frac{1}{3} M_{\rm sat}$.

\textit{Discussion}: we described two general ways to induce half-metal states through a combination of electronic interactions and a finite Zeeman field.
Our work provides new avenue to half-metallic states which have potential applications in future spin-dependent electronics.
Despite our focus on the Hubbard-Heisenberg model on the
triangular lattice,  both of the proposed mechanisms can be easily generalized to other lattice geometries.
It is important to note that in our proposal the Zeeman field sets direction of the spin polarization of conducting electrons in the half-metal state.
This convenient feature makes it different from the recently discussed half-metal SDW state at {\em zero} magnetization \cite{nandkishore2012}
where spin-rotational symmetry breaks spontaneously, resulting in a random selection of the spin polarization axis in the spin space.

Our proposal can also be realized in Kondo lattice systems where the Zeeman field is provided by
exchange couplings between the itinerant electrons and ferromagnetically ordered local moments, see e.g. \cite{martin2008}, as
well as in cold atom systems \cite{partridge2006,zwierlein2006} where it may be easier to achieve the required spin population imbalance.

Magnetic field control of the half-metallic phase may be useful for creating switchable interfaces between half-metal and
noncentrosymmetric superconductor which
have been argued to support Majorana bound states \cite{brouwer2011}.

It is worth noting close physical similarity between our proposal and the previously proposed one-dimensional
`Coulomb drag' setup \cite{pustilnik2006} where role of the lattice is played by the electrons in an active wire interactions
with which gap out one of the spin projections in the passive wire.

Several interesting theoretical questions can be asked regarding the half-metal state. First of all, the metal-insulator transition between a
half-metal and a Mott insulator, found here at $U/t \approx 4.8$, represents Mott transition which is not affected by (gapped) spin fluctuations.
Understanding it in details may lead one to a better characterization of the general Mott transition.
Half metal states are adjacent to many other interesting quantum phases. For example, it was demonstrated \cite{levitov2012} that
$d+id$ chiral superconducting state is the ground state at $3/2$ ($5/8$) filling on triangular (honeycomb) lattice for weak interaction.
If we dope the system with holes while keeping the FS of spin-up electrons perfectly nested by a Zeeman field, the system will
eventually become a half-metal. Understanding how quantum phase transition(s) between these different phases
happen is an interesting question left for future studies.

We would like to thank A. V. Chubukov, L. I. Glazman, M. J. P. Gingras, E. G. Mishchenko, M. E. Raikh and
O. V.  Tchernyshyov for useful discussions.
This work was initiated at the Max-Planck Institute for the Physics of Complex Systems
during the activities of the Advanced Study Group Program on ``Unconventional Magnetism in High Field'', which we would like to thank for hospitality.
The work is supported by NSF through Grant No. DMR-0808842 (O.A.S.) and by NSERC of Canada (Z.H.).
\bibliography{BiblioHF}
\newpage
\appendix{Supplementary Material for "Half-metallic magnetization plateaux" by Zhihao Hao and Oleg A. Starykh}
\label{sec:suppl}
\setcounter{page}{1}
\setcounter{equation}{0}
\setcounter{figure}{0}
\section{Susceptibility}
Susceptibility $\chi_{\sigma}(\mathbf{Q}_a)$ ($a=1,2,3$) of free electrons is given by
\begin{equation}\label{supeqn: sus}
\chi_\sigma(\mathbf{Q}_a)=\frac{1}{A}\int d^2k\frac{n_{\mathbf{k},\sigma}-n_{\mathbf{k}+\mathbf{Q}_a,\sigma}}{\epsilon_{\mathbf{k}}-\epsilon_{\mathbf{k}+\mathbf{Q}_a}}
\end{equation}
where the integration is restricted in the first Brillouin zone of area $A=8\pi^2/\sqrt{3}$. $\epsilon_{\mathbf{k}}=-2\left(\cos k_x+2\cos\frac{k_x}{2}\cos\frac{\sqrt{3}k_y}{2}\right)$ is the dispersion of the non-interacting fermion. Due to six-fold rotational symmetry, $\chi_\sigma(\mathbf{Q}_a)$ is independent of $a$. We consider $\sigma=\uparrow$ and $a=1$.  The particle ($n_{\mathbf{k}}=1$)  and hole ($n_{\mathbf{k}}=0$) states contribute equally to $\chi$. In addition, the integrand is invariant under separate operations $k_x\to -k_x$ and $k_y\to -k_y$. Taking advantage of these symmetries, we compute the leading singular contribution to $\chi$ by writing $\mathbf{k}=(\pi,\pi/\sqrt{3})+\mathbf{q}$ and considering small $q$ limit:
\begin{equation}\label{suseqn:int}
\begin{aligned}
\chi_{\uparrow}(\mathbf{Q}_1)&\approx -\frac{8}{2\sqrt{3}At}\int_{-\Lambda}^{-q_0} dq_x\int_{-\Lambda}^{\sqrt{3}q_x} dq_y \frac{1}{q_xq_y}\\
&\approx -\frac{1}{2\pi^2t}\ln^2\left(\frac{\Lambda}{q_0}\right).\end{aligned}
\end{equation}

For $\sigma=\downarrow$, $\chi_\downarrow$ is computed numerically:
\begin{equation}\label{suseqn:down}
\chi_{\downarrow}(\mathbf{Q}_1)\approx -\frac{0.136}{t}.
\end{equation}
\section{Half-metal state at weak coupling}
\begin{figure}
  \includegraphics[width=0.8\columnwidth]{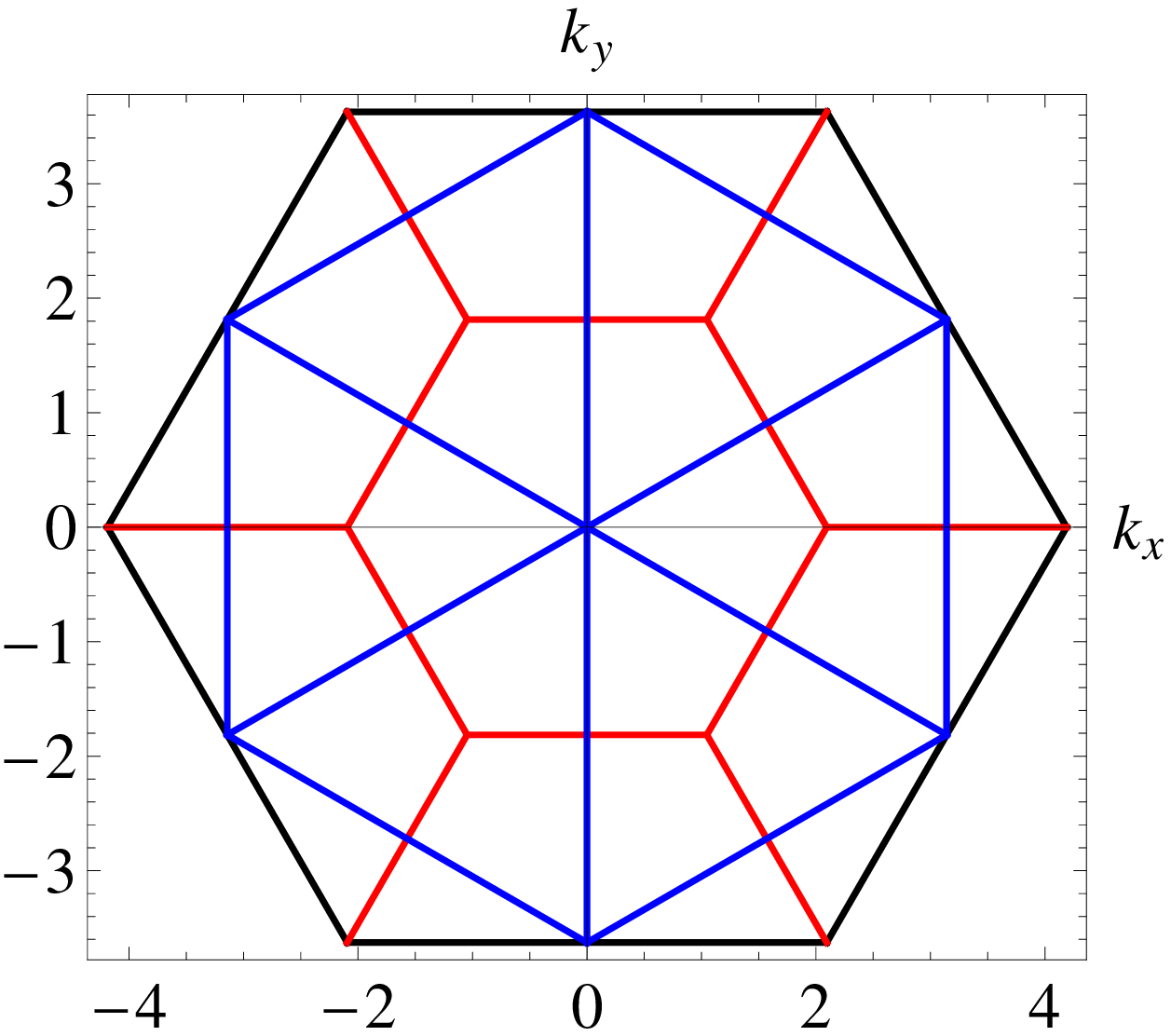}\\
  \caption{The Fermi surface at $y_\uparrow=0$ (the blue lines) under the folded zone scheme. The folded zone (the red hexagons) is four times smaller than the first Brillouin zone of the triangular lattice (the black hexagon). }\label{supfig:folded}
\end{figure}

We consider only spin-up fermions on triangular lattice at $3/4$ filling. The following Hamiltonian is studied:
\begin{equation}\label{supeqn:hami}
H=\sum_{\mathbf{k}}\left((\epsilon_\mathbf{k}-\mu)c_{\mathbf{k} \uparrow}^\dagger c_{\mathbf{k} \uparrow}+
\frac{y}{2}\sum_{a=1}^3(c_{\mathbf{k}\uparrow}^\dagger c_{\mathbf{k}+\mathbf{Q}_q \uparrow}+{\rm h.c})\right).
\end{equation}
The Hamiltonian can be diagonalized and we obtain four bands in the folded zone scheme. In the limit of $y\to 0$, three of the four
bands become $\epsilon_{\mathbf{k}+\mathbf{Q}_a}$ ($a=1,2,3$). These bands touch at three straight lines through the $\Gamma$ point
(Fig. \ref{supfig:folded}). The spectrum at $\mathbf{k}=0$ is $-8$, $-y$, $-y$ and $2y$.
For $y>0$, the lower three bands are $-8$, $-y$ and $-y$.
The Fermi surface is fully gapped. For $y<0$, the Fermi surface become a point at $\mathbf{k}=0$.

\begin{figure}
  \includegraphics[width=0.9\columnwidth]{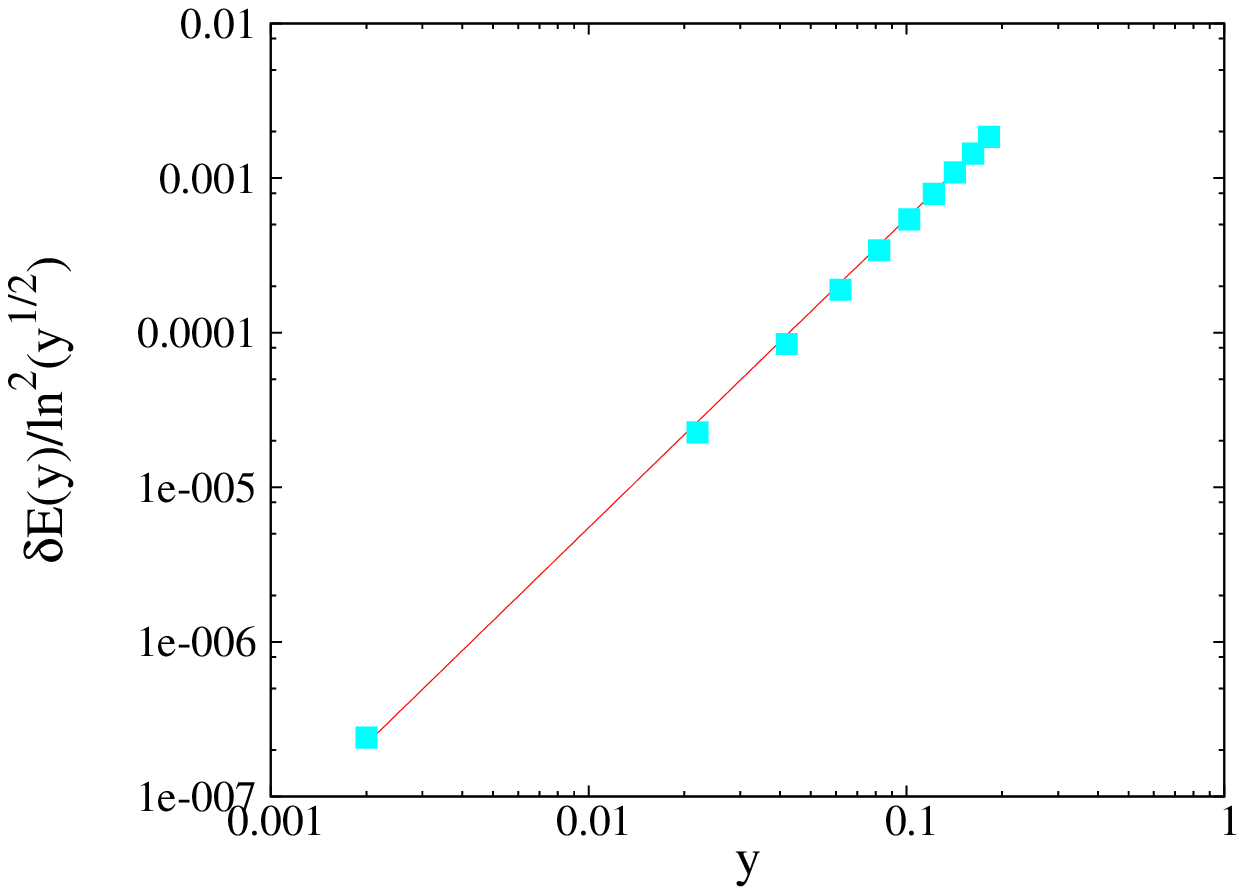}\\
  \caption{$\delta E(y)/\ln^2(y^{1/2})$ (blue squares) as a function of $y$ fitted by function $0.055 y^2$ (the red line). }\label{supfig:diff}
\end{figure}

We define the energy of the Fermi sea to be $E(y)$. At small $|y|$, the band structures of $y>0$ and $y<0$ are
only significantly different around the $\Gamma$ point. $E(y)$ for positive $y$ is lowered by order $|y|$ in an area proportional to
$|y|$ around $\mathbf{k}=0$ since the energies at the van Hove point are functions of $\mathbf{k}^2$ for small $k$.
$\delta E(|y|)\equiv E(-|y|)-E(|y|)$ can then be estimated as proportional to $\chi_{\uparrow}(|y|) y^2\sim\ln^2(|y|^{1/2})y^2$,
as shown by the red line in  Fig. \ref{supfig:diff}. The blue points are obtained by numerically calculating $\delta E(|y|)$ as
a function of momentum $\mathbf{k}$ and summing over fine mesh of $k$-points in the Brillouin zone.

\section{$M=M_{{\rm sat}}/3$ ~half-metal state}
\begin{figure}
  \includegraphics[width=0.95\columnwidth]{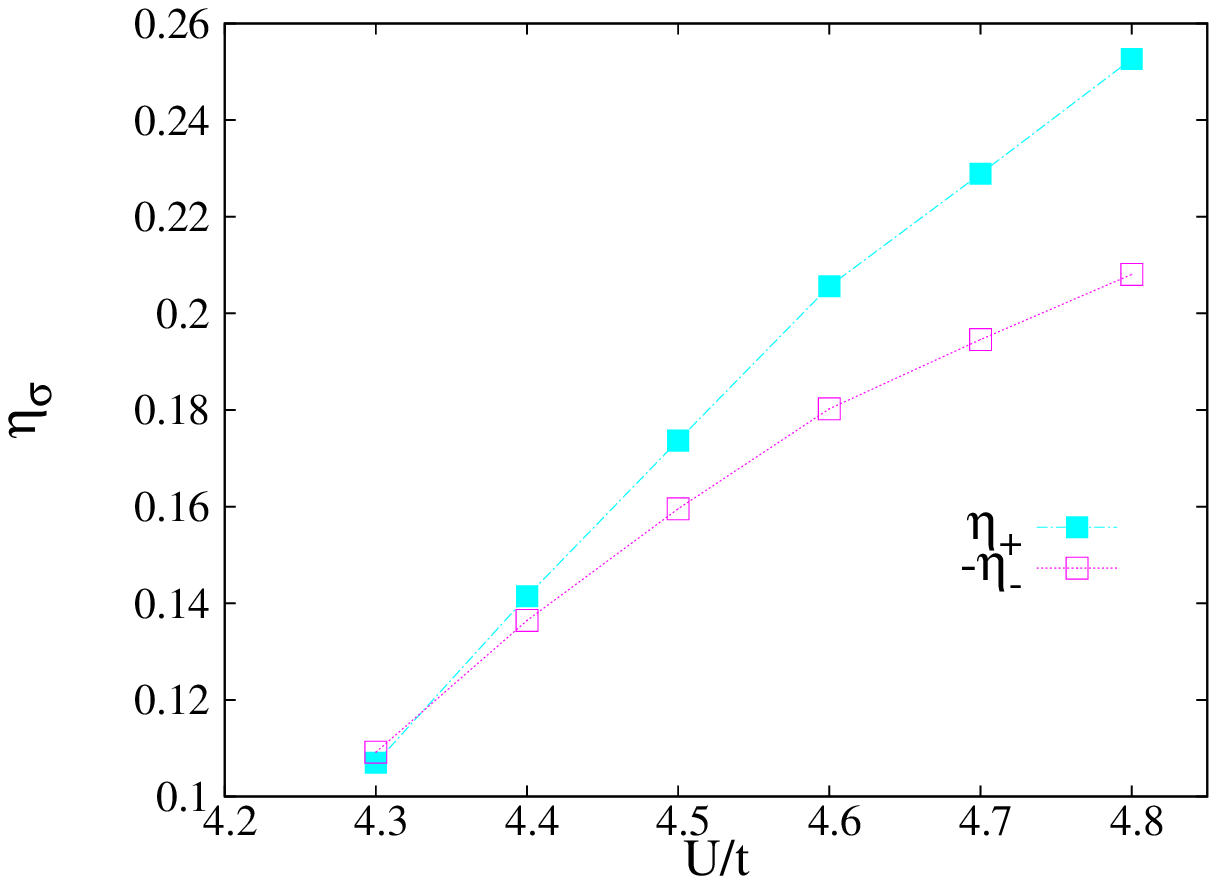}\\
  \caption{$\eta_{\sigma}$ as a function of $U$.}\label{supfig:alleta}
\end{figure}

We use the following mean-field ansatz:
\begin{equation*}
\langle n_{\mathbf{r}\sigma}\rangle=\frac{1}{2}+\frac{\sigma}{6}-2\eta_{\sigma}\cos(\mathbf{Q}\cdot\mathbf{r}).
\end{equation*}
\begin{figure}
  \includegraphics[width=0.8\columnwidth]{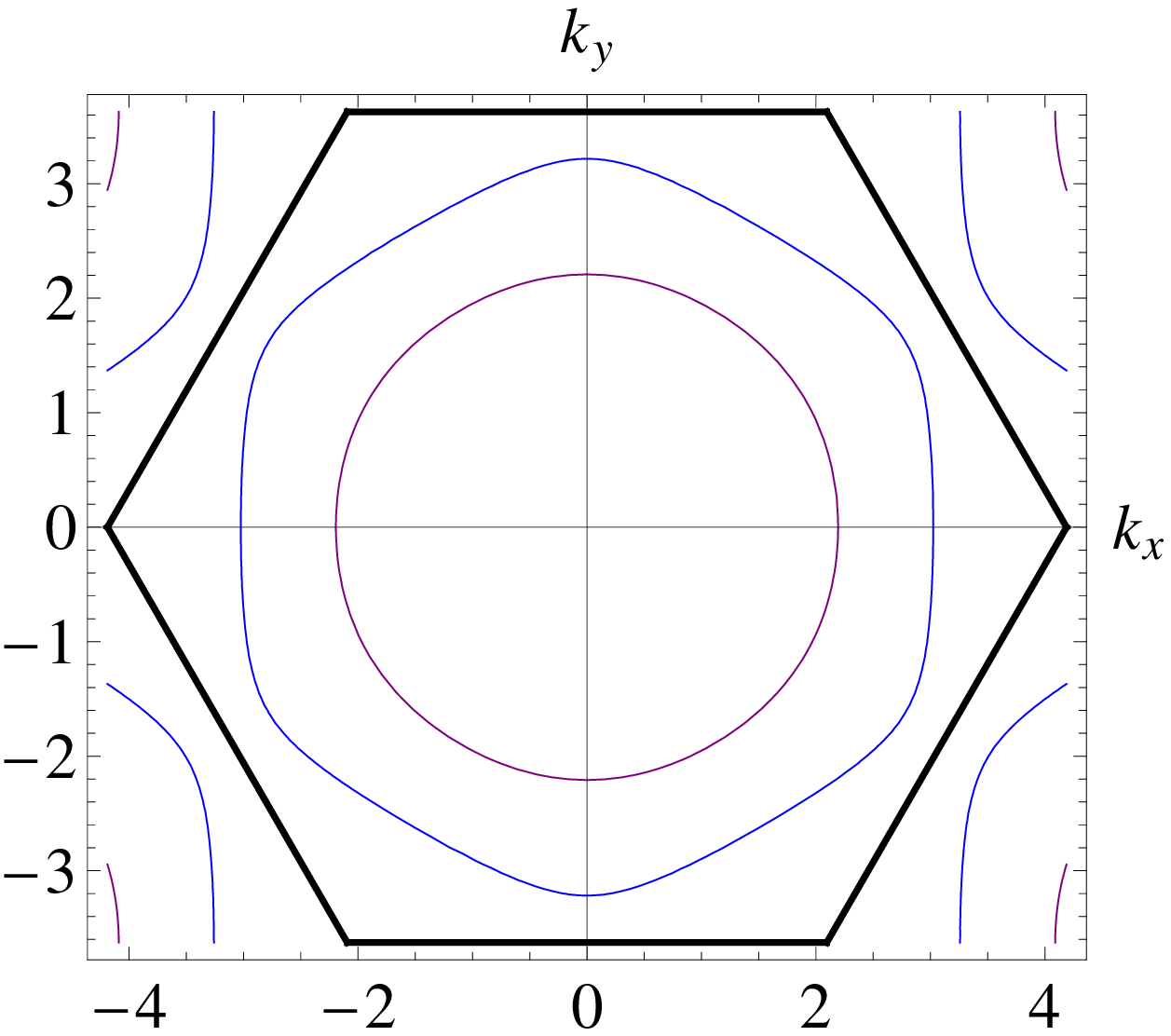}\\
  \caption{The Fermi surfaces of spin-up (blue) and spin-down (purple) electrons at $M=\frac{1}{3} ~M_{\mathrm{sat}}$ for $\eta_{\sigma}=0$. }\label{supfig:onethird}
\end{figure}
The mean-field Hamiltonian can be written as:
\begin{equation}\label{supeqn:mf13}
\begin{aligned}
H=&-\sum_{\langle \mathbf{r}_1\mathbf{r}_2\rangle}(c_{\mathbf{r}_1\sigma}^\dagger c_{\mathbf{r}_2\sigma}+h.c)-\sum_{\mathbf{r}}\left(\mu+\frac{h\sigma}{2}+\frac{U\sigma}{6}\right)c_{\mathbf{r}\sigma}^\dagger c_{\mathbf{r}\sigma}\\
&-2U\sum_{\mathbf{r}}\eta_{-\sigma}\cos(\mathbf{Q}_a\cdot\mathbf{r})c_{\mathbf{r}\sigma}^\dagger c_{\mathbf{r}\sigma}-2NU\eta_{\uparrow}\eta_{\downarrow}
\\&+\frac{5UN}{18}.
\end{aligned}
\end{equation}
The self-consistent equations read:
\begin{subequations}\label{supeqn:self-consist}
\begin{eqnarray}
\eta_{\sigma}&=&-\frac{1}{N}\sum_{\mathbf{r}}\cos(\mathbf{Q}_a\cdot\mathbf{r})\langle c_{\mathbf{r}\sigma}^\dagger c_{\mathbf{r}\sigma}\rangle,\\
1&=&\frac{1}{N}\sum_{\mathbf{r}}c^\dagger_{\mathbf{r}\sigma}c_{\mathbf{r}\sigma}.
\end{eqnarray}
\end{subequations}
These equations are solved numerically.  $\eta_{\sigma}$ become finite for $U>4.3t$ (Fig. \ref{supfig:alleta}).

We compare the energy of the half-metal state with a second mean-field ansatz: a spiral spin-density wave with a uniform magnetization. We use the following notation for order parameters: $\langle S^z_\mathbf{r} \rangle = M$ and $\langle S^+_{\mathbf{Q}}\rangle = m_0 e^{i \mathbf{Q}\cdot \mathbf{r}}$

Define the two component spinor as $\psi_{\mathbf{k}}=(c_{\mathbf{k}+\mathbf{Q}\uparrow},c_{\mathbf{k}\downarrow})^T$, the Hamiltonian can be written as:
\begin{eqnarray}
\label{eqn:spiral_meanfield}
H&=&\sum_{\mathbf{k}}\psi^{\dagger}_{\mathbf{k}}
\Big[(\epsilon_{\mathbf{k}+\mathbf{Q}}+\epsilon_{\mathbf{k}}-\mu)I+(\epsilon_{\mathbf{k}+\mathbf{Q}}-
\epsilon_{\mathbf{k}}-\frac{h}{2}-UM)\sigma_z \nonumber\\
&& - Um_0\sigma_x \Big] \psi_{\mathbf{k}} + N U M^2 + N U m_0^2.
\end{eqnarray}
where $\epsilon_{\mathbf{k}}=-2t(\cos(k_x)+2\cos(k_x/2)\cos(\sqrt{3}k_y/2))$.
The spectrum can be solved:
\begin{equation}\label{eqn:spectrum}
\begin{aligned}
\omega_{\pm}=&\epsilon_{\mathbf{k}+\mathbf{Q}}+\epsilon_{\mathbf{k}}\pm\sqrt{(\epsilon_{\mathbf{k}+\mathbf{Q}}-\epsilon_{\mathbf{k}}-\frac{h}{2}-UM)^2+U^2m_0^2}\\
\equiv &\epsilon_{\mathbf{k}+\mathbf{Q}}+\epsilon_{\mathbf{k}}\pm g_{\mathbf{k}}.
\end{aligned}
\end{equation}
The self consistent equations read:
\begin{equation}
\begin{aligned}
M=&\frac{1}{2N}\sum_{\mathbf{k}}\frac{\epsilon_{\mathbf{k}+\mathbf{Q}}-\epsilon_{\mathbf{k}}-\frac{h}{2}-UM}{g_{\mathbf{k}}}\sum_{\sigma}\sigma\Theta(\mu-\omega_{\sigma}),\\
m_0=&\frac{1}{2N}\sum_{\mathbf{k}}\frac{Um_0}{g_{\mathbf{k}}}\sum_{\sigma}\sigma\Theta(\mu-\omega_{\sigma}),\\
1=&\frac{1}{N}\sum_{\sigma}\Theta(\mu-\omega_{\sigma}).
\end{aligned}
\end{equation}
Here $\Theta(x)=1$ for $x>0$ and $\Theta(x)=0$ for $x<0$.

\begin{figure}
  \includegraphics[width=0.95\columnwidth]{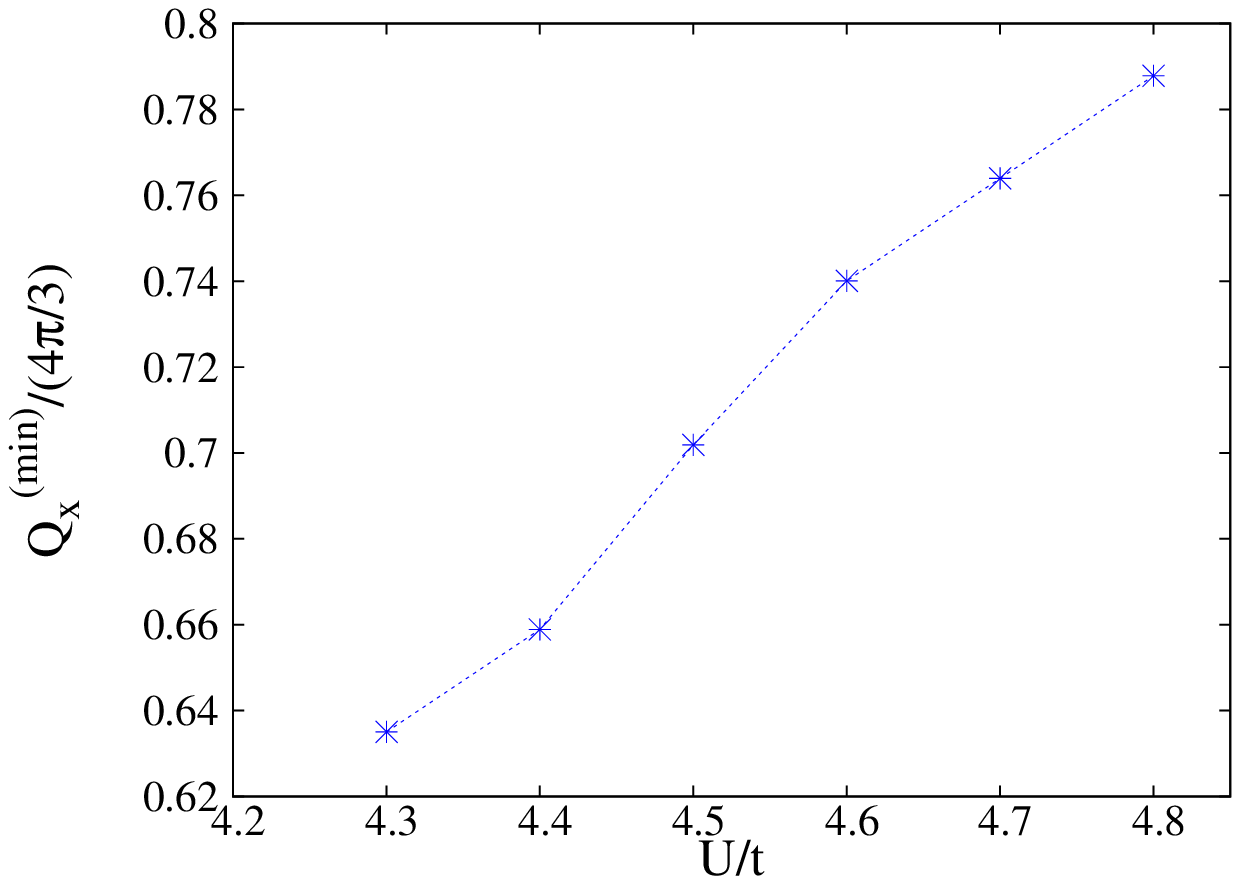}\\
  \caption{$Q_x^{\mathrm{(min)}}/(4\pi/3)$ as a function of $U$. }\label{supfig:qx}
\end{figure}

For fixed $U$ and $h$, we solve the self-consistency equations numerically for different $\mathbf{Q}$'s. 
Focusing on $\mathbf{Q}=(Q_x,0)$, we find the $\mathbf{Q}^{\mathrm{(min)}}$ with lowest mean-field energy. 
$\mathbf{Q}^{(min)}$ is incommensurate in general (Fig. \ref{supfig:qx}). We scan over the two dimensional phase space 
$\mathbf{Q}$ for some $U$'s and confirm that the lowest mean-field energy state is indeed on the $Q_y=0$ axis.

Comparing the mean-field energy of the two states, the half-metal state has lower energy for $U>4.45t$ for ranges of $h$. 
For $U>4.8t$, the half-metal state becomes a band insulator.

\end{document}